\documentstyle[preprint,aps,graphicx]{revtex}

\begin{document}
\draft
\preprint{{\vbox{\hbox {{\bf UCD-HEP-99-24}} \hbox{Dec 1999} }}}

\title {Zee Neutrino Mass Model in a SUSY Framework }
\author{\bf Kingman Cheung$^{1}$ and Otto C. W. Kong$^2$}
\address{$^1$ Department of Physics, University of California, Davis, 
CA 95616 USA \\
$^2$ Institute of Physics, Academia Sinica, Nankang, Taipei TAIWAN 11529}
\maketitle

\begin{abstract}
We study the Zee model of neutrino mass in the framework of $R$-parity
violating supersymmetry.  Within the matter contents of the minimal
supersymmetric standard model, any one of the three right-handed sleptons 
could be  a suitable
candidate for the charged-singlet scalar of the Zee model, and one of
the Higgs doublets provides the extra necessary vacuum expectation
value.  A combination of one bilinear and two trilinear $R$-parity-violating 
couplings then completes the model.  In this framework, we also
discuss other various contributions to neutrino masses and derive the
conditions for the dominance of the contribution from the Zee model, 
and hence maintain the successfully Zee mass texture.  However, this model
within the minimal supersymmetric standard model is shown to be only
marginally feasible.  More general versions of supersymmetrization of
the Zee model are also discussed.  A particularly interesting example 
that has extra Higgs doublets while the slepton, especially the selectron, 
keeps the role of the Zee scalar is illustrated.
\end{abstract}

\section{Introduction}
Within the standard model (SM), the $V-A$ nature of the weak interaction 
dictates a zero mass for all three families of neutrinos. The discovery of 
neutrino 
mass(es) or oscillation(s) will certainly push for new  physics. 
Evidence for neutrino oscillations has been collected in a number of solar
neutrino and atmospheric neutrino experiments.  The most 
impressive results were the recent $\nu_\mu$ neutrino deficit  and the
asymmetric zenith-angle distribution observed by the Super-Kamiokande 
Collaboration \cite{superk}. 

The atmospheric neutrino deficit and zenith-angle distribution
can be explained by the
$\nu_\mu - \nu_\tau$ or $\nu_\mu - \nu_s$ oscillations ($\nu_s$ is a sterile
neutrino that has negligible coupling to $W$ or $Z$ boson and the former
has a slightly better fit).  The oscillation parameters with $\nu_\mu \to
\nu_\tau$ at 90\% C.L. are \cite{superk} 
\begin{displaymath}
\Delta m^2_{\rm atm} \simeq (2 - 6) \times 10^{-3} \; {\rm eV}^2,\;
\sin^2 2 \theta_{\rm atm} \agt 0.85  \;.
\end{displaymath}
On the other hand, the solar neutrino deficit admits more than one solution.
With $\nu_e \to \nu_\tau$ the solutions at 95\% C.L. are \cite{vernon}
\begin{eqnarray}
\mbox{vacuum oscillation:}\;\;\; &&
\Delta m^2_{\rm sol} \simeq (5-8) \times 10^{-11} \; {\rm eV}^2,\;
\sin^2 2 \theta_{\rm sol} \simeq 0.6 - 1.0 \;, \nonumber \\
\mbox{small angle MSW:}\;\;\; &&
\Delta m^2_{\rm sol} \simeq (4-9) \times 10^{-6} \; {\rm eV}^2,\;
\sin^2 2 \theta_{\rm sol} \simeq (3.5 -13)\times 10^{-3}\;,  \nonumber \\
\mbox{large angle MSW:}\;\;\; &&
\Delta m^2_{\rm sol} \simeq (8 - 30) \times 10^{-6} \; {\rm eV}^2,\;
\sin^2 2 \theta_{\rm sol} \simeq 0.4 - 0.8 \;. \nonumber
\end{eqnarray}
The above two neutrino-mass differences can be accommodated by the three 
species
of neutrinos that we know from the SM.  There is also another indication
for neutrino oscillation from the accelerator experiment at the 
Liquid Scintillation Neutrino Detector (LSND) \cite{lsnd}, 
which requires an oscillation of $\nu_\mu$ into another neutrino with 
\begin{displaymath}
\Delta m^2_{\rm LSND} \simeq 0.2 - 2  \; {\rm eV}^2\;,
\sin^2 2 \theta_{\rm LSND} \simeq 0.003 - 0.03 \;.
\end{displaymath}
To as well accommodate this data it requires an additional species of neutrino beyond 
the usual neutrinos.  
Nevertheless, further evidence from the next round of neutrino experiments
is required to confirm the neutrino mass and oscillation.  
As neutrino is favored to be massive it is desirable to understand the 
generation of neutrino masses from physics beyond the SM, especially, to see
if the new physics can give a neutrino mass pattern that can explain the
atmospheric and solar neutrino data, and perhaps the LSND as well.

An economical way to generate small neutrino masses with a 
phenomenologically favorable texture is given by the Zee model
\cite{zee,fg,jmst}, which generates masses via one-loop diagrams.
The model consists of a charged gauge singlet scalar $h^{\mbox{-}}$, 
the Zee scalar, which couples to lepton doublets 
$\psi_{\!\scriptscriptstyle Lj}$ via the interaction
\begin{equation} \label{zcp}
f^{ij} \left( \psi^\alpha_{\!\scriptscriptstyle Li} {\cal C} 
\psi^\beta_{\!\scriptscriptstyle Lj} \right ) 
\epsilon_{\!\scriptscriptstyle \alpha\beta}\; 
h^{\scriptscriptstyle -} \;,
\end{equation}
where $\alpha,\beta$ are the SU(2) indices, $i,j$ are the generation indices, 
${\cal C}$ is the charge-conjugation matrix, 
and $f^{ij}$ are Yukawa couplings antisymmetric in $i$ and $j$.  
Another ingredient of the Zee model is an extra Higgs doublet (in addition to 
the one that gives masses to charged leptons) that develops a 
vacuum expectation value (VEV) and thus provides mass mixing between
the charged Higgs boson and the Zee scalar boson. The corresponding
coupling, together with the $f^{ij}$'s, enforces lepton number violation.
The one-loop mechanism for the Zee model can be found in Fig.~\ref{fig1}.

A recent analysis by Frampton and Glashow \cite{fg} (see also 
Ref.\cite{jmst}) showed that the Zee mass 
matrix of the following texture 
\begin{equation} 
\label{zee-m}
 \left( \begin{array}{ccc}
0        & m_{e\mu} & m_{e\tau} \\
m_{e\mu} & 0        & \epsilon  \\
m_{e\tau}& \epsilon & 0 
         \end{array}          \right ) \; ,
\end{equation}
where $\epsilon$ is small compared with $m_{e\mu}$ and $m_{e\tau}$,
is able to provide a compatible mass pattern that explains the atmospheric
and solar neutrino data. The generic Zee model guarantees the vanishing 
of the diagonal elements, while the suppression of the $m_{\mu\tau}$ entry,
here denoted by the small parameter $\epsilon$, has to be 
otherwise enforced.  Moreover, $m_{e\mu} \sim m_{e\tau}$ is required to
give the maximal mixing solution for the atmospheric neutrinos.
We shall describe the features in more detail in the next section.

So far the Zee model is not embedded into any grand unified theories or
supersymmetric models.  Here we analyze the embedding of the Zee model 
into the minimal supersymmetric standard model (MSSM) with minimal 
extensions, namely, the $R$-parity violation.  The right-handed sleptons 
in SUSY have the right quantum number to play the role of the charged 
Zee scalar.  The $R$-parity-violating $\lambda$-type couplings ($\lambda LLE$) 
could provide the terms in Eq.(\ref{zcp}). It is also easy to see that 
the $R$-parity-violating bilinear $\mu$-type couplings ($\mu_i 
L H_{\scriptscriptstyle 2}$) 
would allow the second Higgs doublet $H_{\scriptscriptstyle 2}$ in
SUSY to be the second ingredient of the Zee model.
So far so good. However, the SUSY framework dictates extra 
contributions to neutrino masses, which deviate from the texture 
of the Zee mass matrix of Eq. (\ref{zee-m}). 
The major objective of this paper is to address the 
feasibility of the embedding and to determine under what 
conditions could one make a supersymmetric Zee model within the 
$R$-parity-violating SUSY framework while retaining the successful flavor of 
the former.  We will also discuss briefly more generic versions of 
supersymmetric Zee model.
There is also a study of Zee mass matrix within the framework of 
gauge-mediated SUSY breaking with the messenger field as the Zee singlet
\cite{haba}.

In $R$-parity-violating SUSY, there are three 
other sources for neutrino masses, 
in addition to the Zee model contribution.  They are
(i) the tree-level mixing with the higgsinos and gauginos, (ii) the 
one-loop diagram that involves the usual mass mixing between the left-handed 
and right-handed sleptons proportional to $m_{\scriptscriptstyle \ell}\, 
(A^{\!\scriptscriptstyle E}_{\scriptscriptstyle \ell} - \mu \tan\beta)$,
and (iii) the one-loop diagram that again involves the mixing between 
the left-handed and right-handed sleptons but this time via the 
$\lambda$ and $\mu_i$ couplings.
\footnote{
There is one other type of contribution from a gaugino-sneutrino loop
with neutrino-antineutrino mass splitting. The latter could be a result
of $R$-parity violation. This was discussed in Ref. \cite{gh}.
This contribution depends on the $R$-parity violating $B$ terms and basically 
enters in the same entries in the neutrino mass matrix as the corresponding
$\mu_i$ terms, and thus supplementing the latter. We will neglect this type
of contribution in this paper.
}
The first two contributions have been considered extensively in literature
\cite{previous}, but the last one is identified here for the first time.  
Also, we are the first one to identify the
Zee model contribution to the neutrino mass in the SUSY framework. 
Furthermore, we will obtain the conditions for the 
Zee model contribution to dominate over the contributions in (i) and (ii).
The contribution in (iii) can actually preserve the texture of the mass 
matrix of Eq. (\ref{zee-m}).

There are complications in choosing a flavor basis when $R$ parity
is broken.  Actually, the form and structure of the lepton mass matrices under 
the coexistence of bilinear and trilinear $R$-parity-violating couplings 
are basis dependent. One has to be particularly
careful with a consistent choice of flavor basis.  Here we adopt the 
single-VEV parametrization \cite{svp} that provides an efficient
framework for our study.  The most important point to note here is that
this parametrization implies a choice of
flavor basis under which all three sneutrinos have no VEV, without any
input assumptions. All $R$-parity-violating couplings introduced below are
to be interpreted under this basis choice.

The organization of the paper is as follows. In the next section, 
we describe the texture of the Zee mass matrix in Eq. (\ref{zee-m}).
In Sec. III, we calculate entries in the $3\times 3$ neutrino mass matrix 
from all the sources listed above.
In Sec. IV, we derive the conditions for the contributions from the Zee model
and from (iii) above to be dominant. Section V is devoted to discussions on
more general versions of supersymmetrization of the Zee model.
We conclude in Sec. VI.

\section{Zee Mass  Matrix}
Here we briefly describe the basic features of the Zee mass matrix, as given in
Eq.(\ref{zee-m}).  We first take $\epsilon=0$. The matrix can be 
diagonalized by the following transformation
\begin{equation}
\left( \begin{array}{c}
         \nu_{\!\scriptscriptstyle Le} \\
         \nu_{\!\scriptscriptstyle L\mu} \\
         \nu_{\!\scriptscriptstyle L\tau} \end{array} \right )
=
\left( \begin{array}{ccc}
   \frac{1}{\sqrt{2}} &  \frac{1}{\sqrt{2}} &  0 \\
   \frac{m_{e\mu}}{\sqrt{2} m} & \frac{-m_{e\mu}}{\sqrt{2} m} &
   \frac{- m_{e\tau}}{m}  \\
   \frac{m_{e\tau}}{\sqrt{2} m} & \frac{-m_{e\tau}}{\sqrt{2} m} &
   \frac{m_{e\mu}}{m}  \end{array} \right ) \;
\left( \begin{array}{c}
         \nu_{\!\scriptscriptstyle L_1} \\
         \nu_{\!\scriptscriptstyle L_2} \\
         \nu_{\!\scriptscriptstyle L_3} \end{array} \right ) \;,
\end{equation}
with the eigenvalues $m, -m, 0$ for $\nu_{\!\scriptscriptstyle L_1},
\nu_{\!\scriptscriptstyle L_2},\nu_{\!\scriptscriptstyle L_3}$, respectively,
and $m=\sqrt{m_{e\mu}^2 + m_{e\tau}^2}$. Hence, the two massive states
form a Dirac pair. The atmospheric mass-squared difference
$\Delta m^2_{\rm atm} \simeq  3\times 10^{-3} {\rm eV}^2$, is to be 
identified with 
$m^2=m_{e\mu}^2 + m_{e\tau}^2$.  The transition probabilities for 
$\nu_{\!\scriptscriptstyle L_\mu}$ are
\begin{eqnarray}
P_{\nu_{\scriptscriptstyle L_\mu} \to \nu_{\scriptscriptstyle L_e} } &=& 0 
\;, \nonumber \\
P_{\nu_{\scriptscriptstyle L_\mu} \to \nu_{\scriptscriptstyle L_\tau} } 
&=& 4 \left( \frac{m_{e\mu} m_{e\tau} }
{ m_{e\mu}^2 + m_{e\tau}^2 } \right )^2 \sin^2  \left(
\frac{ (m_{e\mu}^2 + m_{e\tau}^2) L }{4 E} \right )  \;. \nonumber 
\end{eqnarray}
If $m_{e\mu}\simeq m_{e\tau}$, then $\sin^2 2\theta_{\rm atm} \simeq 1$.  This
mixing angle is exactly what is required in the atmospheric neutrino data.
The neutrino mass matrix texture with $\epsilon=0$ can be called the zeroth 
order Zee texture. It is the first thing to aim at in our supersymmetric 
model discussions in the next section.

If we choose a nonzero $\epsilon$, but keep $\epsilon \ll m_{e\mu, e\tau}$.
Then after diagonalizing the matrix we have the following eigenvalues
\begin{eqnarray}
m_{\nu\scriptscriptstyle 1} &=& \sqrt{m_{e\mu}^2 + m_{e\tau}^2} + 
  \epsilon \frac{m_{e\mu} m_{e\tau}}{ m_{e\mu}^2 + m_{e\tau}^2 }\;, \nonumber\\
m_{\nu\scriptscriptstyle 2} &=& -\sqrt{m_{e\mu}^2 + m_{e\tau}^2} + 
  \epsilon \frac{m_{e\mu} m_{e\tau}}{ m_{e\mu}^2 + m_{e\tau}^2 }\;,\nonumber \\
m_{\nu\scriptscriptstyle 3} &=& 
-2 \epsilon \frac{m_{e\mu} m_{e\tau}}{ m_{e\mu}^2 + m_{e\tau}^2 }
\;.  \nonumber 
\end{eqnarray}
The mass-square difference between $m^2_{\nu\scriptscriptstyle 1}$ 
and $m^2_{\nu\scriptscriptstyle 2}$ can be fitted to the
solar neutrino mass.  For instance, one can take the large angle 
Mikheyev-Smirnov-Wolfenstein (MSW) solution and requires
\begin{displaymath}
4 \epsilon \frac{m_{e\mu} m_{e\tau}}{\sqrt{ m_{e\mu}^2 + m_{e\tau}^2 }} =
 \Delta m^2_{\rm sol} \simeq 2 \times 10^{-5} \;{\rm eV}^2 \; ,
\end{displaymath}
giving 
\begin{displaymath}
\frac{\epsilon}{m_{e\mu}} \sim 5\times 10^{-3} \;,
\end{displaymath}
where we have used $m_{e\mu} \simeq m_{e\tau}$.

We will see below that in our supersymmetric model the couplings that are
required to generate the zeroth order Zee texture also give 
rise to other contributions, which have to be kept subdominating in
order to maintain the texture and hence the favor of the Zee model. 
Even though these extra contributions might not be identified as the 
same entries as the ${\epsilon}$ parameter of Eq. (\ref{zee-m}), i.e.,
appear in the diagonal entries instead, they 
could still play the same role as to give a phenomenologically viable 
first order result for a modified Zee matrix. 
Hence, we will not commit 
ourselves to the first order Zee matrix as given in Eq.(\ref{zee-m}), 
but only to its zeroth order form, namely with ${\epsilon}=0$. The first 
order perturbation is then allowed to come in through any matrix entry.
It will split the mass square degeneracy of the Dirac pair similar
to the ${\epsilon}$ case above.
For example, if  the first order perturbation is
given by a  $\epsilon_{\scriptscriptstyle d}$ appearing at the $m_{ee}$ entry,
the resulting mass eigenvalues are modified to
\begin{eqnarray}
m_{\nu\scriptscriptstyle 1} &=& \sqrt{m_{e\mu}^2 + m_{e\tau}^2} + 
  \frac{\epsilon_{\scriptscriptstyle d}}{2} \;, \nonumber\\
m_{\nu\scriptscriptstyle 2} &=& -\sqrt{m_{e\mu}^2 + m_{e\tau}^2} + 
  \frac{\epsilon_{\scriptscriptstyle d}}{2} \;, \nonumber\\
m_{\nu\scriptscriptstyle 3} &=& 0 \;.  \nonumber 
\end{eqnarray}
The mass-square difference between $m^2_{\nu\scriptscriptstyle 1}$ 
and $m^2_{\nu\scriptscriptstyle 2}$ can then be fitted to the solar 
neutrino data and we obtain 
$\epsilon_{\scriptscriptstyle d}/m_{e\mu} \sim 5 \times 10^{-3}$, 
the same as above.  If, on the other hand, $\epsilon_{\scriptscriptstyle d}$ 
appears at the $m_{\mu\mu}$ or
$m_{\tau\tau}$ entry, the solar neutrino data can still be 
fitted and $\epsilon_{\scriptscriptstyle d}/m_{e\mu} \sim 1\times 10^{-2}$ 
is required. Once the zeroth order Zee texture for the
atmospheric neutrino is satisfied, it is straightforward to further 
impose the above condition for the solar neutrino.

\section{Neutrino Mass Matrix}
First consider the superpotential as given by
\begin{equation}
W = \epsilon_{\!\scriptscriptstyle \alpha\beta} \Biggr \{
Y_{ij}^{\!\scriptscriptstyle U} Q^\alpha_i H_{\!\scriptscriptstyle 2}^\beta 
U_j^c 
+ Y_{ij}^{\!\scriptscriptstyle D}
 Q_i^\alpha H_{\!\scriptscriptstyle 1}^\beta D^c_j
+ Y^{\!\scriptscriptstyle E}_{ij} L_i^\alpha H_{\!\scriptscriptstyle 1}^\beta 
E^c_j \nonumber \\
+ \mu H_{\!\scriptscriptstyle 1}^\alpha H_{\!\scriptscriptstyle 2}^\beta 
+ \lambda_{ijk} L_i^\alpha L_j^\beta E^c_k
+\mu_i L_i^\alpha H_{\!\scriptscriptstyle 2}^\beta  
\Biggr\}\; ,
\end{equation}
where $\epsilon_{\scriptscriptstyle 12}=-\epsilon_{\scriptscriptstyle 21}=-1$,
$i,j=1,2,3$ are the generation indices. 
$H_{\scriptscriptstyle 1}=(h^0_{\scriptscriptstyle 1}, 
h^{\mbox{-}}_{\scriptscriptstyle 1}), H_{\scriptscriptstyle 2} = 
( h^+_{\scriptscriptstyle 2}, h^0_{\scriptscriptstyle 2})$.  In the above 
equation, $Q,L,U^c,D^c,E^c,H_{\scriptscriptstyle 1}$, and 
$H_{\!\scriptscriptstyle 2}$ denote the quark doublet, 
lepton doublet, up-quark singlet, down-quark singlet, lepton singlet,
and the two Higgs
doublet superfields.  Here we allow only the $R$-parity violation through
the terms $LLE^c$ and $LH_{\!\scriptscriptstyle 2}$ with coefficients 
$\lambda_{ijk}$ 
(antisymmetric in $i,j$) and $\mu_i$, respectively. The other 
$R$-parity-violating
couplings are dropped as they are certainly beyond the minimal framework
needed for  embedding the Zee model. The soft SUSY breaking terms that 
are relevant to our study are
\[
 (Y^{\!\scriptscriptstyle E} \!A^{\!\scriptscriptstyle E})_{ij} 
\tilde{L}^\alpha_i H_{\!\scriptscriptstyle 1}^\beta \tilde{E}^c_j 
+ (\lambda A^{\!\lambda})_{ijk}\tilde{L}^\alpha_i \tilde{L}^\beta_j 
\tilde{E}^c_k
+ \mu B H_{\!\scriptscriptstyle 1}^\alpha H_{\!\scriptscriptstyle 2}^\beta \;. 
\]
Actually, the $(\lambda A^{\!\lambda})$ terms do not contribute because
our choice of basis eliminates the VEV's for $\tilde{L}_i$'s. 
This simplifies the analysis without lose of generality. 
We adopt the single-VEV parametrization, which uses the $L_i$ basis 
such that the charged-lepton Yukawa matrix 
$Y^{\!\scriptscriptstyle E}$ is diagonal. The whole $(Y^{\!\scriptscriptstyle 
E} \!A^{\!\scriptscriptstyle E})$ term
will be taken as predominantly diagonal, namely, $(Y^{\!\scriptscriptstyle E} 
\!A^{\!\scriptscriptstyle E})_{ij} \approx Y^{\!\scriptscriptstyle E}_i
A^{\!\scriptscriptstyle E}_{i} \delta_{ij} \,\mbox{(no sum)}$. This is just the
common practice of suppressing off-diagonal $A$ terms, favored by 
flavor-changing neutral-current constraints. 

The tree-level mixing among the higgsinos, gauginos, and neutrinos gives rise
to a $7\times 7$ neutral fermion mass matrix $\cal{M_N}$:
\begin{equation}
\label{77}
\cal{M_N} = 
\left (\begin{array}{cccc|ccc}
M_{\scriptscriptstyle 1} & 0 & g'v_{\scriptscriptstyle 2}/2 
& -g' v_{\scriptscriptstyle 1}/2 & 0 & 0 & 0  \\
0   & M_{\scriptscriptstyle 2} & -gv_{\scriptscriptstyle 2}/2 
& g v_{\scriptscriptstyle 1}/2 & 0 & 0 & 0  \\
g'v_{\scriptscriptstyle 2}/2 & -g v_{\scriptscriptstyle 2}/2 & 0   
& -\mu & -\mu_{\scriptscriptstyle 1} & -\mu_{\scriptscriptstyle 2} 
& -\mu_{\scriptscriptstyle 3} \\
-g'v_{\scriptscriptstyle 1}/2 & g v_{\scriptscriptstyle 1}/2 & -\mu 
&   0 & 0 & 0 & 0\\
\hline
0 & 0 & -\mu_1     & 0 & (m_\nu^0)_{\!\scriptscriptstyle 1\!1} 
& (m_\nu^0)_{\!\scriptscriptstyle 1\!2} & (m_\nu^0)_{\!\scriptscriptstyle 1\!3}
\\
0 & 0 &  -\mu_{\!\scriptscriptstyle 2} & 0  & (m_\nu^0)_{\!\scriptscriptstyle 
2\!1} 
& (m_\nu^0)_{\!\scriptscriptstyle 22} & (m_\nu^0)_{\!\scriptscriptstyle 23} \\
0 & 0 & -\mu_{\!\scriptscriptstyle 3} & 0 & (m_\nu^0)_{\!\scriptscriptstyle 31}
& (m_\nu^0)_{\!\scriptscriptstyle 32} & (m_\nu^0)_{\!\scriptscriptstyle 33} \\
 \end{array}
\right )\; ,
\end{equation}
whose basis is $(-i\tilde{B}, -i\tilde{W}, 
\tilde{h}_{\scriptscriptstyle 2}^0, \tilde{h}_{\scriptscriptstyle 1}^0, 
\nu_{\!\scriptscriptstyle L_e},\nu_{\!\scriptscriptstyle L_\mu },
\nu_{\!\scriptscriptstyle  L_\tau}) $. Each of the 
charged-lepton states deviates from its physical state as a result
of its mixing with higgsino-gaugino through the corresponding 
$\mu_i$ term\cite{svp}. However, we are interested only in a
region of the parameter space where the concerned deviations are
negligible, as also discussed in Ref.\cite{otto}. Hence, we are 
effectively in the basis of the physical charged-lepton states, as indicated.
In the above $7\times 7$ 
matrix, the whole lower-right $3\times 3$ block $(m_\nu^0)$ is zero at 
tree level.  They are induced via one-loop contributions to be discussed below.
One-loop contributions to the other zero entries are neglected.
We can write the mass matrix in the form of block submatrices:
\begin{equation}
{\cal M_N} = \left( \begin{array}{c|c}
              {\cal M} & \xi^{\!\scriptscriptstyle  T} \\
\hline
              \xi & m_\nu^0 \end{array}  \right ) \;,
\end{equation}
where $\cal{M}$ is the upper-left $4\times 4$ neutralino mass matrix, 
$\xi$ is the $3\times 4$ block, and $m_\nu^0$ is the lower-right 
$3\times 3$ neutrino block in the $7\times 7$ matrix.  
The resulting neutrino mass matrix after block diagonalization is given by
\begin{equation} \label{mnu}
(m_\nu) = - \xi {\cal M}^{\mbox{-}1} \xi^{\!\scriptscriptstyle  T} + (m_\nu^0)
 \;.
\end{equation}
The first term here corresponds to tree level contributions, which are, 
however, see-saw suppressed.

Before going into our best scenario analysis, we will sketch how the
couplings, $\lambda_{ijk}$'s and $\mu_i$'s, lead to the neutrino mass 
terms. While some of them have been studied in literature, others are 
identified here for the first time.  We do this from the perspectives of
the supersymmetric Zee model, but the results are quite general.

Our minimalistic strategy says that a  $\lambda_{ijk}$ or a $\mu_i$ should 
be taken as zero unless it is needed for the Zee mechanism to generate 
the neutrino mass terms $m_{e\mu}$ and $m_{e\tau}$. Readers who find the 
extensive use of unspecified indices in the following discussions 
difficult to follow are suggested to match them with the results for the 
explicit examples that we will list below.  We identify the following four
neutrino-mass generation mechanisms.

(i) {\it Zee mechanism}.
We show in Fig.~\ref{fig1}, the two Zee diagrams for the one-loop neutrino 
mass terms.  The right-handed slepton $\tilde{\ell}_{\!\scriptscriptstyle R_k}$
is identified as the charged-singlet boson of the Zee model, and its 
coupling to lepton fields has the correct antisymmetric generation 
indices: see Eq. (\ref{zcp}).  To complete the diagram the 
charged Higgs boson $h_{\scriptscriptstyle 1}^{\mbox{-}}$ from the Higgs 
doublet $H_{\scriptscriptstyle 1}$ is on 
the other side of the loop and a 
$\tilde{\ell}_{\!\scriptscriptstyle R_k}$-$h_1^{\mbox{-}}$ mixing
is needed at the top of the loop.
Such a mixing is provided by a  $F$ term of $L_k$:
$\mu_k m_{\!\scriptscriptstyle \ell_k} h_{\scriptscriptstyle 1}^{\mbox{-}} 
\tilde{\ell}^*_{\!\scriptscriptstyle R_k} \langle h_{\scriptscriptstyle 2}^0
\rangle / \langle h_{\scriptscriptstyle 1}^0 \rangle$,
where $h_{\scriptscriptstyle 2}^0$ takes on its VEV, for a nonzero 
$\mu_k$.  Thus, the neutrino mass term $(m_\nu^0)_{ij}$ has a
\begin{equation}
\mu_k m_{\!\scriptscriptstyle \ell_{k}}\lambda_{ijk}
(m_{\!\scriptscriptstyle \ell_j}^2 -m_{\!\scriptscriptstyle \ell_i}^2)
\end{equation}
dependence, where $m_{\!\scriptscriptstyle \ell_i}$'s
are the charged lepton masses. 

(ii) {\it $LR$ slepton mass mixing}.
Another, well-studied, type of contributions comes from the one-loop diagram 
with two $\lambda$-coupling vertices and the usual
$(A^{\!\scriptscriptstyle E}-\mu\tan\beta)$-type 
$LR$ slepton mixing. Neglecting the
off-diagonal entries in $A^{\!\scriptscriptstyle E}$, 
the contribution to $(m_\nu^0)_{ij}$ with the pair $\lambda_{ilk}$ and 
$\lambda_{jkl}$ is proportional to 
\begin{equation}
\label{Aloop}
 \big[ \;(A^{\!\scriptscriptstyle E}_k - \mu \tan\!{\beta})
+ (1-\delta_{kl}) (A^{\!\scriptscriptstyle E}_l - \mu \tan\!{\beta})\;\big]\;
m_{\!\scriptscriptstyle \ell_k} m_{\!\scriptscriptstyle \ell_l} 
 \lambda_{ilk}\lambda_{jkl}  \;.
\end{equation}
Only $\lambda_{ijk}$'s with all distinct indices (e.g. 
$\lambda_{\scriptscriptstyle 1\!23}$)
fail to give contributions of this kind on its own. A nonzero
$\lambda_{ikk}$ contributes to the diagonal $(m_\nu^0)_{ii}$. 
An illustration for the term is given in Fig.~\ref{fig2}. With any two nonzero
$\lambda_{ijk}$'s, this kind of contributions to the $(m_\nu^0)$ entries, 
in particular the diagonal ones, cannot be avoided. 

(iii) {\it $LR$ slepton mass mixing via $R$-parity violating couplings}.
This contribution is identified here for the first time.  
While the contributions to generation mixing in the usual 
$(A^{\!\scriptscriptstyle E}-\mu\tan\beta)$-type $LR$ slepton mixing via the 
off-diagonal entries in $A^{\!\scriptscriptstyle E}$ are expected to be small,
there is another independent source of 
generation mixing in the $LR$ slepton-mass mixing, 
which may not follow the rule.  
The latter comes from a  $F$ term of $L_i$: $\mu_i\lambda_{ijk} 
\tilde{\ell}_{\!\scriptscriptstyle L_j} \tilde{\ell}^*_{\!\scriptscriptstyle 
R_k} 
\langle h_{\scriptscriptstyle 2}^0 \rangle$, where 
$h_{\scriptscriptstyle 2}^0$ takes on the VEV.  
This is similar to the $\tilde{\ell}_{\!\scriptscriptstyle R_k}
\!\mbox{-}h_{\scriptscriptstyle 1}^{\mbox{-}}$ mixing in the Zee model,
except that this time we have a $\lambda$-type coupling instead of
the $R$-parity-conserving Yukawa
coupling.  This newly identified source of mixing
results in constraints on the $\mu_i \lambda_{ijk}$ products,
which is an interesting subject of lepton-slepton phenomenology studies.  

With a specific choice of a set of nonzero 
$\mu_i$'s and $\lambda$'s, this type of mixing gives rise to 
the off-diagonal $(m_\nu^0)_{ij}$ terms only and, therefore,
of particular interest to our perspectives of Zee model.
Taking the pair $\lambda_{ilk}$ and $\lambda_{jhl}$ for the fermion
vertices and a  
$F$ term of $L_g$ providing a coupling for the scalar vertex
in the presence of a $\mu_g$ and a $\lambda_{ghk}$ (see Fig.~\ref{fig3}),
a $(m_\nu^0)_{ij}$ term is generated and proportional to 
\begin{equation}
\label{ghk}
\mu_g m_{\!\scriptscriptstyle \ell_l}\lambda_{ghk} \lambda_{ilk} \lambda_{jhl}
 \; .
\end{equation}
The proliferation of indices here is certainly difficult to keep track of. 
When we allow only a single nonzero $\lambda$ at a time, the only 
contribution comes from $\lambda_{ijj}$ but not from those with distinct 
indices.  Suppose we have nonzero $\lambda_{ijj}$  and $\mu_j$, 
they then give a contribution to the off-diagonal $(m_\nu^0)_{ij}$ with a
\[
\mu_j m_{\!\scriptscriptstyle \ell_j} \lambda_{ijj}^3 
\]
dependence, which is obtained from expression (\ref{ghk}) 
through the substitution $h=i$ and $g=k=l=j$.  
It is easy to see that for a minimal set of nonzero $\mu_i$ and 
$\lambda_{ijk}$ required to generate the zeroth order Zee texture,
this minimal set also contributes to the
same neutrino mass terms via the new mechanism identified here.
Hence, they are desirable from the perspectives of keeping the Zee mass 
matrix texture.

(iv) {\it Tree-level mixing}.
Through gaugino-higgsino mixings, 
nonzero $\mu_i$'s give tree-level see-saw type contributions to 
$(m_\nu)_{ij}$ proportional to  $\mu_i \mu_j$, i.e., through the first
term in Eq.(\ref{mnu}) instead of the second. With the contribution put in
explicitly, Eq.(\ref{mnu}) then gives
\begin{equation} \label{mnu2}
(m_\nu)_{ij} =  - \,\frac{v^2 \cos^2\!\beta 
\;( g^2 M_{\scriptscriptstyle 1} + g^{'2} M_{\scriptscriptstyle 2} )}
{2 \mu \;[2 \mu M_{\scriptscriptstyle 1} M_{\scriptscriptstyle 2} 
- v^2 \sin\!\beta \cos\!\beta \;(g^2 M_{\scriptscriptstyle 1} 
+ g^{'2}M_{\scriptscriptstyle 2} ) ]}\;\mu_i \mu_j
 + (m_\nu^0)_{ij}\; .
\end{equation}
A diagonal $(m_\nu)_{kk}$ term is 
always present for a nonzero $\mu_k$ as needed in the Zee mechanism. 
This contribution has no charged lepton mass dependence.  To 
eliminate these tree-level terms requires either very stringent constraints
on the parameter space or extra Higgs superfields 
beyond the MSSM spectrum. We will see that this is a major difficulty of 
the present MSSM formulation of supersymmetric Zee model. 

From the above discussions, we conclude that a minimal set of $R$-parity
violating couplings needed to give the zeroth order Zee matrix is the 
following :
\[
\{  \quad
\lambda_{{\scriptscriptstyle 12}\,k}\;,\quad
\lambda_{{\scriptscriptstyle 13}\,k}\;,\quad
\mu_k \quad \}\; .
\]
As at least one of the two $\lambda$'s has the form $\lambda_{ikk}$
($\equiv -\lambda_{kik}$), 
all types of contributions that have been discussed above are there. 
We want to make the contribution from the Zee mechanism dominate over other
contributions, or at least to make the diagonal mass entries to $(m_\nu)$
subdominant.  This necessarily requires subdomination of the 
contributions from the tree-level see-saw mechanism and from the 
$(A^{\!\scriptscriptstyle E} - \mu \tan\!{\beta})$-type $LR$ slepton mixing. 
So, it is the Zee mechanism and the newly identified mechanism, 
which involve the interplay between the bilinear $\mu_i$ and trilinear 
$\lambda_{ijk}$ $R$-parity-violating couplings, that are required to be 
the dominating ones.

We will discuss below two illustrative scenarios
(1) $\lambda_{\scriptscriptstyle 1\!2\!1}$, 
$\lambda_{\scriptscriptstyle 1\!3\!1}$, and $\mu_{\scriptscriptstyle 1}$ and
(2) $\lambda_{\scriptscriptstyle 1\!23}$, 
$\lambda_{\scriptscriptstyle 1\!33}$,  and $\mu_{\scriptscriptstyle 3}$.
After the $7\times 7$ matrix ${\cal M}_{\cal N}$ is block diagonalized,
the resulting $3\times 3$ neutrino mass matrix $(m_\nu)$ of Eqs.(\ref{mnu})
or (\ref{mnu2}) is obtained in each of these scenarios.

{\it Scenario 1: $\lambda_{\scriptscriptstyle 1\!2\!1}$, 
$\lambda_{\scriptscriptstyle 1\!3\!1}$, and $\mu_{\scriptscriptstyle 1}$}.
The resulting neutrino mass matrix for scenario 1 is given by
\small \begin{equation}
\label{scen1}
(m_\nu) = 
\left( \begin{array}{ccc}
C_{\scriptscriptstyle 1} \, \mu_{\scriptscriptstyle 1}^2 
\;\;  & \;\; 
 C_{\scriptscriptstyle 2}\, m_e \, m_\mu^2 \,
\mu_{\scriptscriptstyle 1} \lambda_{\scriptscriptstyle 1\!2\!1}
+ C_{\scriptscriptstyle 3} \, m_e \,\mu_{\scriptscriptstyle 1} 
\lambda_{\scriptscriptstyle 1\!2\!1} 
\;\;  & \;\;
C_{\scriptscriptstyle 2} \, m_e \, m_\tau^2 \,
\mu_{\scriptscriptstyle 1} \lambda_{\scriptscriptstyle 1\!3\!1}
 + C_{\scriptscriptstyle 3} \, m_e \,\mu_{\scriptscriptstyle 1} 
\lambda_{\scriptscriptstyle 1\!3\!1}
\\
   & C_{\scriptscriptstyle 4} \, m_e^2 \,\lambda_{\scriptscriptstyle 1\!2\!1}^2
  & 2 C_{\scriptscriptstyle 4}\, m_e^2  \,\lambda_{\scriptscriptstyle 1\!2\!1}
\lambda_{\scriptscriptstyle 1\!3\!1} \\
& & C_{\scriptscriptstyle 4}\, m_e^2  \,\lambda_{\scriptscriptstyle 1\!3\!1}^2 
\end{array} \right )\; ,
\end{equation} \normalsize
which is symmetric and we only write down the upper triangle.  The $C_i$'s are
given by
\begin{eqnarray}
C_{\scriptscriptstyle 1} &=& - \,\frac{v^2 \cos^2\!\beta \;
( g^2 M_{\scriptscriptstyle 1} + g^{'2} M_{\scriptscriptstyle 2} )}
{2 \mu \;[2 \mu M_{\scriptscriptstyle 1} M_{\scriptscriptstyle 2} 
- v^2 \sin\!\beta \cos\!\beta\; (g^2 M_{\scriptscriptstyle 1} 
+ g^{'2}M_{\scriptscriptstyle 2} ) ]} \;, \nonumber \\
C_{\scriptscriptstyle 2} &=& \frac{-1}{16\pi^2} 
\frac{\sqrt{2}\tan\!\beta} {v \cos\!\beta}\;  
f( M_{h_{\scriptscriptstyle 1}^{\mbox{-}}}^2, M_{\tilde{e}_R}^2 )
\; , \nonumber \\
C_{\scriptscriptstyle 3} &=& \frac{1}{16\pi^2} 
\frac{v\sin\!\beta}{\sqrt{2}} \left[ \lambda_{\scriptscriptstyle 1\!2\!1}^2
  \; f( M_{\tilde{\mu}_L}^2, M_{\tilde{e}_R}^2 )  
+  \lambda_{\scriptscriptstyle 1\!3\!1}^2  
\; f( M_{\tilde{\tau}_L}^2,M_{\tilde{e}_R}^2 ) \right ]
\; , \nonumber \\
C_{\scriptscriptstyle 4} &=& - \frac{1}{16\pi^2}\; 
( A^{\!\scriptscriptstyle E}_e - \mu \tan\!\beta) \;
         f(M^2_{\tilde{e}_L},M^2_{\tilde{e}_R}) \;, 
\end{eqnarray}
where 
\[
f(x,y) = \frac{1}{x-y} \; \log\left( \frac{y}{x} \right ) \;. 
\]

{\it Scenario 2: $\lambda_{\scriptscriptstyle 1\!23}$, 
$\lambda_{\scriptscriptstyle 1\!33}$,  and $\mu_{\scriptscriptstyle 3}$}.
For scenario 2, the neutrino mass matrix is given by
\small \begin{equation}
\label{scen2}
(m_\nu) = 
\left( \begin{array}{ccc}
C'_{\scriptscriptstyle 4}\, m_\tau^2\, \lambda_{\scriptscriptstyle 1\!33}^2
\;\; & \;\;
C'_{\scriptscriptstyle 2} \, m_\tau \, m_\mu^2 \,
\mu_{\scriptscriptstyle 3} \lambda_{\scriptscriptstyle 1\!23} 
+ C_{\scriptscriptstyle 5}\;
 m_\tau \, \mu_{\scriptscriptstyle 3} \lambda_{\scriptscriptstyle 1\!23}
 \lambda_{\scriptscriptstyle 1\!33}^2
\;\; & \;\;
   C'_{\scriptscriptstyle 2} \, m_\tau^3 \, 
\mu_{\scriptscriptstyle 3} \lambda_{\scriptscriptstyle 1\!33} 
+ C_{\scriptscriptstyle 5}\;
 m_\tau \, \mu_{\scriptscriptstyle 3}  \lambda_{\scriptscriptstyle 1\!33}^3
 \\    & 0 & 0 \\
   & & C_{\scriptscriptstyle 1} \;\mu_{\scriptscriptstyle 3}^2  
\end{array} \right )\; ,
\end{equation} \normalsize
where
\begin{eqnarray}
C'_{\scriptscriptstyle 4} &=& - \frac{1}{16\pi^2} \; 
( A^{\!\scriptscriptstyle E}_\tau - \mu \tan\!\beta) 
\;f(M^2_{\tilde{\tau}_L},M^2_{\tilde{\tau}_R}) 
 \;, \nonumber \\
 C'_{\scriptscriptstyle 2} &=&\frac{-1}{16\pi^2} 
\frac{\sqrt{2}\tan\!\beta} {v \cos\!\beta}\;  
f( M_{h_{\scriptscriptstyle 1}^{\mbox{-}}}^2, M_{\tilde{\tau}_R}^2 )
\; , \nonumber \\
C_{\scriptscriptstyle 5} &=&- \frac{1}{16\pi^2}  \frac{v\sin\!\beta}{\sqrt{2}}
 f( M_{\tilde{e}_L}^2, M_{\tilde{\tau}_R}^2 )\;.
\end{eqnarray} 
In the above, we have neglected terms suppressed by $m_e/m_\mu$ or 
$m_e/m_\tau$.  There is also another scenario, with  
\{$\lambda_{\scriptscriptstyle 1\!22}, \lambda_{\scriptscriptstyle 1\!32},
\mu_{\scriptscriptstyle 2}$\}, which is very similar to this scenario 2.

\section{conditions for maintaining the Zee mass texture}

In order to maintain the zeroth order Zee texture as discussed in
Sec. II, we need $m_{e\mu}$ and $m_{e\tau}$ to dominate over the other
entries. Moreover, we need $m_{e\mu}\sim m_{e\tau}\sim\sqrt{\Delta
M^2_{\rm atm}} (\sim 5\times 10^{-11}\;\mbox{GeV})$.  Here we give an
estimate of the required conditions on the model parameters, for a
chosen minimal set of $R$-parity-violating couplings
\{$\lambda_{{\scriptscriptstyle 12}k},\lambda_{{\scriptscriptstyle
13}k,}\;\mu_k$; (with a specific $k$)\}.  Since we are interested only in the
absolute value of each term and so we will drop negative signs
wherever feasible. We will look at each matrix entry in
Eqs. (\ref{scen1}) and (\ref{scen2}) carefully.

{\it Scenario 1}.
Requiring the tree-level gaugino-higgsino mixing contribution to $m_{ee}$
in Eq. (\ref{scen1}) to be negligible compared to $m_{e\mu}$, it gives
\begin{equation}
\label{tree1}
{\mu_{\scriptscriptstyle 1}^2}\, \cos^2\!\!\beta \ll 
{\mu^2} M_{\scriptscriptstyle 1}\,(1\times 10^{-14}\, {\rm GeV}^{-1}) \;,
\end{equation} 
in which we have assumed 
$M_{\scriptscriptstyle 1} \approx 0.5M_{\scriptscriptstyle 2}$.
The condition is rather stringent that requires either very small 
$\mu_{\scriptscriptstyle 1}$, 
$\mu_{\scriptscriptstyle 1}\cos\!\beta\ll 10^{-4}\,\mbox{GeV}$ at 
$M_{\scriptscriptstyle 1},\mu \sim O(100) \,\mbox{GeV}$, or particularly
large gaugino mass(es). As pointed out in Ref.\cite{svp}, the dependence on
$\tan\!\beta$ is very important here. The $\cos\!\beta$ goes from order
one to $\sim 0.02$ in the domain of large $\tan\!\beta$.

Requiring the $(A^{\!\scriptscriptstyle E}_k - \mu \tan\!\beta)$ $LR$ slepton 
mixing contribution to be much smaller than $m_{e\mu}$, we have 
\begin{equation} \label{lamll}
\lambda_{\scriptscriptstyle 1\!2\!1}^2, \lambda_{\scriptscriptstyle 1\!3\!1}^2
\ll \frac{\mbox{max}(M^2_{\tilde{e}_L},M^2_{\tilde{e}_R})}
{(A^{\scriptscriptstyle E}_e - \mu \tan\!\beta)}\;
(3\times 10^{-2}\,\mbox{GeV}^{-1})   \;,
\end{equation}
where we have used
\begin{equation}
\left[  f(M^2_{\tilde{e}_L},M^2_{\tilde{e}_R}) \right]^{-1}
\sim\; \mbox{max}(M^2_{\tilde{e}_L},M^2_{\tilde{e}_R})\; .
\end{equation}
The constraint in Eq. (\ref{lamll}) is obviously very weak. In fact, 
it can certainly be neglected,
especially when other phenomenological constraints \cite{jmst,czee} on
$\lambda$'s [as effective Zee couplings $f_{ij}$ of Eq. (\ref{zcp})]
are taken into consideration. The relevant constraint here is given as
\begin{equation} \label{mudy}
\frac{\lambda_{\scriptscriptstyle 1\!2\!1}^2}{M^2_{\tilde{e}_R}}
\leq 10^{-8}\,\mbox{GeV}^{-2} \;,
\end{equation}
from the tree-level Zee-scalar mediated $\mu$ decay \cite{mueg}. 
The upper bound on $\lambda_{\scriptscriptstyle 1\!2\!1}$
is hence no better than $0.01$ for $M_{\tilde{e}_R}$ at 
$100\,\mbox{GeV}$. The corresponding constraint on 
$\lambda_{\scriptscriptstyle 1\!3\!1}$ from $\tau$ decay is definitely weaker,
which has no relevance here as we will see below. Hence, the
suppression needed for $m_{\mu\mu}$, $m_{\mu\tau}$, and $m_{\tau\tau}$
of Eq. (\ref{scen1}) is easy to obtain.

The remaining question is if one can still generate the right (order of) 
$m_{e\mu}$ and $m_{e\tau}$ when $\mu_{\scriptscriptstyle 1}$,
$\lambda_{\scriptscriptstyle 1\!2\!1}$ and 
$\lambda_{\scriptscriptstyle 1\!3\!1}$
satisfy the above constraints. Let us first look at the
$m_{e\mu}$ entry. From Eq.(\ref{scen1}), $m_{e\mu}$ has two contributions. 
The first one (the one with a $C_{\scriptscriptstyle 2}$ dependence) is from 
the authentic Zee mechanism. 
For this contribution to give the right $m_{e\mu}$ value, it requires
\begin{equation} \label{zm12}
m_{e\mu}\;\sim\;
\frac{\mu_{\scriptscriptstyle 1} \lambda_{\scriptscriptstyle 1\!2\!1}}
{\cos^2\!\!\beta}\;
\frac{1}{\mbox{max}( M_{h_{\scriptscriptstyle 1}^{\mbox{-}}}^2, 
M_{\tilde{e}_R}^2 )} \;
(2\times 10^{-10}\,\mbox{GeV}^{2})\; 
\sim \;(5\times 10^{-11}\, {\rm GeV}) \; ,
\end{equation}
or
\begin{equation}
(\mu_{\scriptscriptstyle 1} \cos\!\beta) \;\lambda_{\scriptscriptstyle 1\!2\!1}
\sim \;{\cos^3\!\!\beta}\;{\mbox{max}
( M_{h_{\scriptscriptstyle 1}^{\mbox{-}}}^2, 
M_{\tilde{e}_R}^2 )}\; (0.25 \,\mbox{GeV}^{-1})\;.
\end{equation}
The right-hand side above cannot be much smaller than $0.01\,\mbox{GeV}$, even 
at the more favorable case of very large $\tan\!\beta$. 
Given the above constraints in Eqs. (\ref{tree1}) and (\ref{mudy}), 
this is obviously unrealistic.  Even though the corresponding
contribution (with a $C_2$ dependence) to $m_{e\tau}$ has a 
$m_\tau^2/m_\mu^2$ enhancement and depends
on $\lambda_{\scriptscriptstyle 1\!3\!1}$ instead of
$\lambda_{\scriptscriptstyle 1\!2\!1}$, it is not much better than 
$m_{e\mu}$.

For the second contribution (the one with a 
$C_{\scriptscriptstyle 3}$ dependence),
to give the right $m_{e\mu}$ value, it requires
\begin{equation}
\label{26}
m_{e\mu}\;\sim\; \mu_{\scriptscriptstyle 1} 
\left[ \frac{\lambda_{\scriptscriptstyle 1\!2\!1}^3}
  {\mbox{max}( M_{\tilde{\mu}_L}^2, M_{\tilde{e}_R}^2 )}  
+\frac{\lambda_{\scriptscriptstyle 1\!3\!1}^2  
         \lambda_{\scriptscriptstyle 1\!2\!1}}
{\mbox{max}( M_{\tilde{\tau}_L}^2,M_{\tilde{e}_R}^2 )} \right ] 
 (5\times 10^{-4}\,\mbox{GeV}^{2})\;
\sim \;(5\times 10^{-11}\, {\rm GeV}) \; .
\end{equation}
A naive comparison with Eq. (\ref{zm12}) above illustrates one important fact.
Assuming a common scale for the  scalar masses, the $\lambda$ coupling(s) 
only have to be larger than $10^{-3}$ for this second contribution to be 
larger than the first one. From Eqs. (\ref{lamll}) and (\ref{mudy}), 
such $\lambda$'s are easily admissible. 
With $\lambda_{\scriptscriptstyle 1\!2\!1}
\approx \lambda_{\scriptscriptstyle 1\!3\!1}$ the corresponding contribution to
$m_{e\tau}$ has the same form as $m_{e\mu}$, with the interchange of 
$\lambda_{\scriptscriptstyle 1\!2\!1}$ with 
$\lambda_{\scriptscriptstyle 1\!3\!1}$, and thus is
of a similar value. The condition in Eq. (\ref{26}) then becomes
\begin{equation}
(\mu_{\scriptscriptstyle 1}  \cos\!\beta)\; 
\lambda_{{\scriptscriptstyle 1}i{\scriptscriptstyle 1}}  \; 
\left(\frac{\lambda_{{\scriptscriptstyle 1}i{\scriptscriptstyle 1}}^2}
{M_{\tilde{e}_R}^2 }\right)\;  \;
\sim \;\cos\!\beta \;(5\times 10^{-8}\, {\rm GeV}^{-1}) \; ,
\end{equation}
where we have taken $M_{\tilde{e}_R}$ to be the dominating mass 
among the scalars. The latter choice corresponds to the optimal case
because smaller scalar masses help reducing the size 
of the $\lambda_{1i1}$ needed, while on the other hand, in Eq. (\ref{mudy})
larger $M_{\tilde{e}_R}$ relaxes the constraint on 
$\lambda_{{\scriptscriptstyle 1}i{\scriptscriptstyle 1}}^2$.
With Eqs. (\ref{tree1}) and (\ref{mudy}) 
taken into consideration, the result 
ends up actually no better than the best (large $\tan\!\beta$) case of 
Eq.(\ref{zm12}) above.

{\it Scenario 2}.
Here we follow our above analysis for this more interesting scenario.
Requiring the tree-level gaugino-higgsino mixing contribution 
to be well below $m_{e\mu}$ gives
\begin{equation} \label{mtt}
{\mu_{\scriptscriptstyle 3}^2}\, \cos^2\!\!\beta \ll 
{\mu^2} M_{\scriptscriptstyle 1}\,(1\times 10^{-14}\, {\rm GeV}^{-1}) \;.
\end{equation} 
This is basically the same as in scenario 1, though it corresponds to 
$m_{\tau\tau}$ instead.  

For the $(A^{\scriptscriptstyle E}_k - \mu \tan\!\beta)$ $LR$ slepton mixing
contribution to be much smaller than $m_{e\mu}$, we have 
\begin{equation} \label{2a}
\lambda_{\scriptscriptstyle 1\!33}^2
\ll \frac{\mbox{Max}(M^2_{\tilde{\tau}_L},M^2_{\tilde{\tau}_R})}
{(A^{\scriptscriptstyle E}_\tau - \mu \tan\!\beta)}\;
 (2.5 \times 10^{-9} \,\mbox{GeV}^{-1})\;.
\end{equation}
This corresponds to $m_{ee}$. 
It tells us that $\lambda_{\scriptscriptstyle 1\!33}$ can hardly be much 
larger than $10^{-3}$.
On the other hand, $\lambda_{\scriptscriptstyle 1\!23}$ is constrained
differently because  
it does not contribute to this type of neutrino mass term.
The constraint that corresponds to Eq.(\ref{mudy}), however, becomes 
\begin{equation} \label{mudy1}
\frac{\lambda_{\scriptscriptstyle 1\!23}^2}{M^2_{\tilde{\tau}_R}}
\leq 10^{-8}\,\mbox{GeV}^{-2} \;,
\end{equation}
which tells us that $\lambda_{\scriptscriptstyle 1\!23}$ can be as large as
order of $0.01$ for scalar masses of order of $O(100)$ GeV. 

Again both $m_{e\mu}$ and $m_{e\tau}$ have two terms.  Let us look at
$m_{e\mu}$ first.
For the first term in $m_{e\mu}$ (the one with a $C'_{\scriptscriptstyle 2}$
dependence) in Eq. (\ref{scen2}) to give the required value of
atmospheric neutrino mass, we need 
\begin{equation} \label{2zm12}
m_{e\mu}\;\sim\;
\frac{\mu_{\scriptscriptstyle 3} \lambda_{\scriptscriptstyle 1\!23}}
{\cos^2\!\!\beta}\;
\frac{1}{\mbox{max}( M_{h_{\scriptscriptstyle 1}^{\mbox{-}}}^2, 
M_{\tilde{\tau}_R}^2 )} \;
(7\times 10^{-7}\,\mbox{GeV}^{2})\; 
\sim \;(5\times 10^{-11}\, {\rm GeV})
\end{equation}
or
\begin{equation}
(\mu_{\scriptscriptstyle 3} \cos\!\beta) \;\lambda_{\scriptscriptstyle 1\!23}
\sim \;{\cos^3\!\!\beta}\;{\mbox{max}( 
M_{h_{\scriptscriptstyle 1}^{\mbox{-}}}^2, 
M_{\tilde{\tau}_R}^2 )}\; (7\times 10^{-5} \,\mbox{GeV}^{-1})\;.
\end{equation}
This result looks relatively promising. If we take $\cos\!\beta=0.02$,
all the involved scalar masses at $100\,\mbox{GeV}$ and 
$\lambda_{\scriptscriptstyle 1\!23}$ at the corresponding limiting $0.01$ 
value, $\mu_{\scriptscriptstyle 3} \cos\!\beta$ has to be at $5.6\times 10^{-4}
\,\mbox{GeV}$ to fit the requirement. This means pushing for larger
$M_{\scriptscriptstyle 1}$ (and  $M_{\scriptscriptstyle 2}$) and $\mu$ values
but may not be ruled out. 

What about the corresponding first term in $m_{e\tau}$ 
entry? The term has a 
$\lambda_{\scriptscriptstyle 1\!33}$ dependence in the place of 
$\lambda_{\scriptscriptstyle 1\!23}$ with  an extra enhancement of 
$m_\tau^2/m_\mu^2$, in comparison to $m_{e\mu}$. That is to say, requiring
$m_{e\mu}\approx m_{e\tau}$ gives, in this case,
\begin{equation}
\label{29}
\lambda_{\scriptscriptstyle 1\!33} \approx \frac{m_\mu^2}{m_\tau^2}\;
\lambda_{\scriptscriptstyle 1\!23} \; .
\end{equation}
This gives a small $\lambda_{\scriptscriptstyle 1\!33}$ easily satisfying
Eq. (\ref{2a}). The small $\lambda_{\scriptscriptstyle 1\!33}$ also suppresses
the second terms in both $m_{e\mu}$ and $m_{e\tau}$, the
$C_{\scriptscriptstyle 5}$ dependent terms in Eq. (\ref{scen2}).
Note that the above equation represents a kind of fine-tuned relation
between the two couplings $\lambda_{\scriptscriptstyle 1\!33}$
and $\lambda_{\scriptscriptstyle 1\!23}$. More precisely, the value of 
$\lambda_{\scriptscriptstyle 1\!33}\,{m_\tau^2}$ has to be
within a factor of $1.8$ of that of 
$\lambda_{\scriptscriptstyle 1\!23}\,{m_\mu^2}$ in order
to fit $\sin\!\!^22\theta_{\rm atm}$. This feature is inherited
directly from the original Zee model, as discussed in Ref.\cite{jmst}.
Nevertheless, it is difficult to motivate this relation from a theoretical 
point of view.  Phenomenologically,
the relation implies that $\lambda_{\scriptscriptstyle 1\!33}$ is two
orders of magnitude smaller than $\lambda_{\scriptscriptstyle 1\!23}$,
which indicates a strongly inverted hierarchy against the familiar flavor
structure among quarks and charged leptons.  Since the current experimental
bounds from the rare processes, such as $\mu,\tau \to e \gamma$, showed the 
usual hierarchical trend down the families, the relation in Eq. (\ref{29})
says that once the constraints on the $\lambda_{\scriptscriptstyle 1\!23}$ 
are satisfied, $\lambda_{\scriptscriptstyle 1\!33}$ should be automatically
safe.  This justifies our above statement that the $\tau$-decay constraint
analogous to Eqs. (\ref{mudy}) and (\ref{mudy1}) have no 
relevancy here.
Conversely, if $\lambda_{\scriptscriptstyle 1\!33}$ contributions to
some rare processes are identified in the near future, it would spell
trouble for the SUSY Zee model discussed here.

Finally, we comment on whether it is feasible to have an alternative 
situation in which the second  ($C_{\scriptscriptstyle 5}$ 
dependent) terms in $m_{e\mu}$ and $m_{e\tau}$ dominate 
over the first ($C'_{\scriptscriptstyle 2}$ dependent)  terms.
The comparison between these two types of contributions is similar 
to that of scenario 1, as can be easily seen by comparing terms in 
Eqs. (\ref{scen1}) and (\ref{scen2}). As in scenario 1, we need to push 
$\lambda_{\scriptscriptstyle 1\!33}$ to the order of $0.01$. 
This at the same time requires either a particularly large 
$M_{\tilde{\tau}_L}$ or some fine-tuned cancellation 
between $A^{\scriptscriptstyle E}_\tau$ and $\mu \tan\!\beta$ 
in order to fulfill the condition in Eq. (\ref{2a}).  Thus, it is unlikely
to have the second terms of $m_{e\mu}$ and $m_{e\tau}$ dominant over the
first terms.

To produce the neutrino mass matrix beyond the zeroth order Zee texture, 
the subdominating first-order contributions are required to be 
substantially smaller in order to fit the solar neutrino data. Here, 
it is obvious that it is difficult to further suppress the tree level 
gaugino-higgsino mixing contribution to $m_{\tau\tau}$, which makes it 
even more difficult to get the scenario to work. Explicitly, the requirement
for the solar neutrino is
\begin{equation} 
{\mu_{\scriptscriptstyle 3}^2}\, \cos^2\!\!\beta \sim 
{\mu^2} M_{\scriptscriptstyle 1}\,(1\times 10^{-16}\, {\rm GeV}^{-1}) \;,
\end{equation} 
following directly from the result given in Sec. II [cf. Eq. (\ref{mtt})].

\section{more general versions of supersymmetric Zee model}
We have discussed in detail the minimalistic embedding of the Zee model
into the minimal supersymmetric standard model. The conditions for maintaining
the Zee neutrino mass matrix texture is extremely stringent, if not impossible.
Here we discuss some more general versions of supersymmetrization of the 
Zee model.

\begin{figure}[th]
\centering
\includegraphics[width=5.5in]{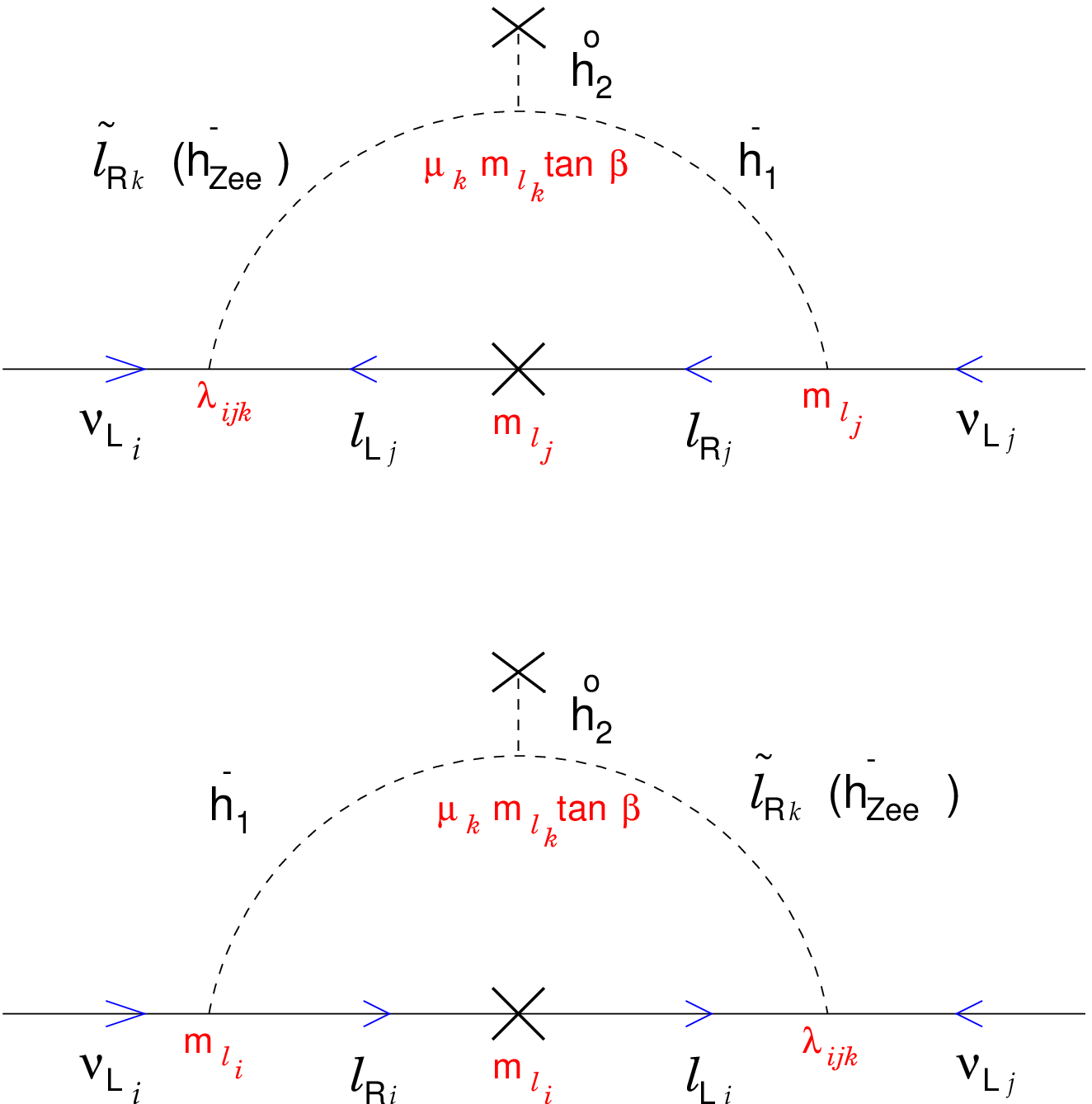}
\vskip1in
\caption{\label{fig1}
\small 
The two Feynman diagrams for the Zee mechanism in the $R$-parity violating
SUSY framework. The original charged singlet boson 
$h_{\rm Zee}^{\mbox{-}}$ 
of the Zee model is shown in parentheses.}
\end{figure}

\begin{figure}[th]
\centering
\includegraphics[width=5.5in]{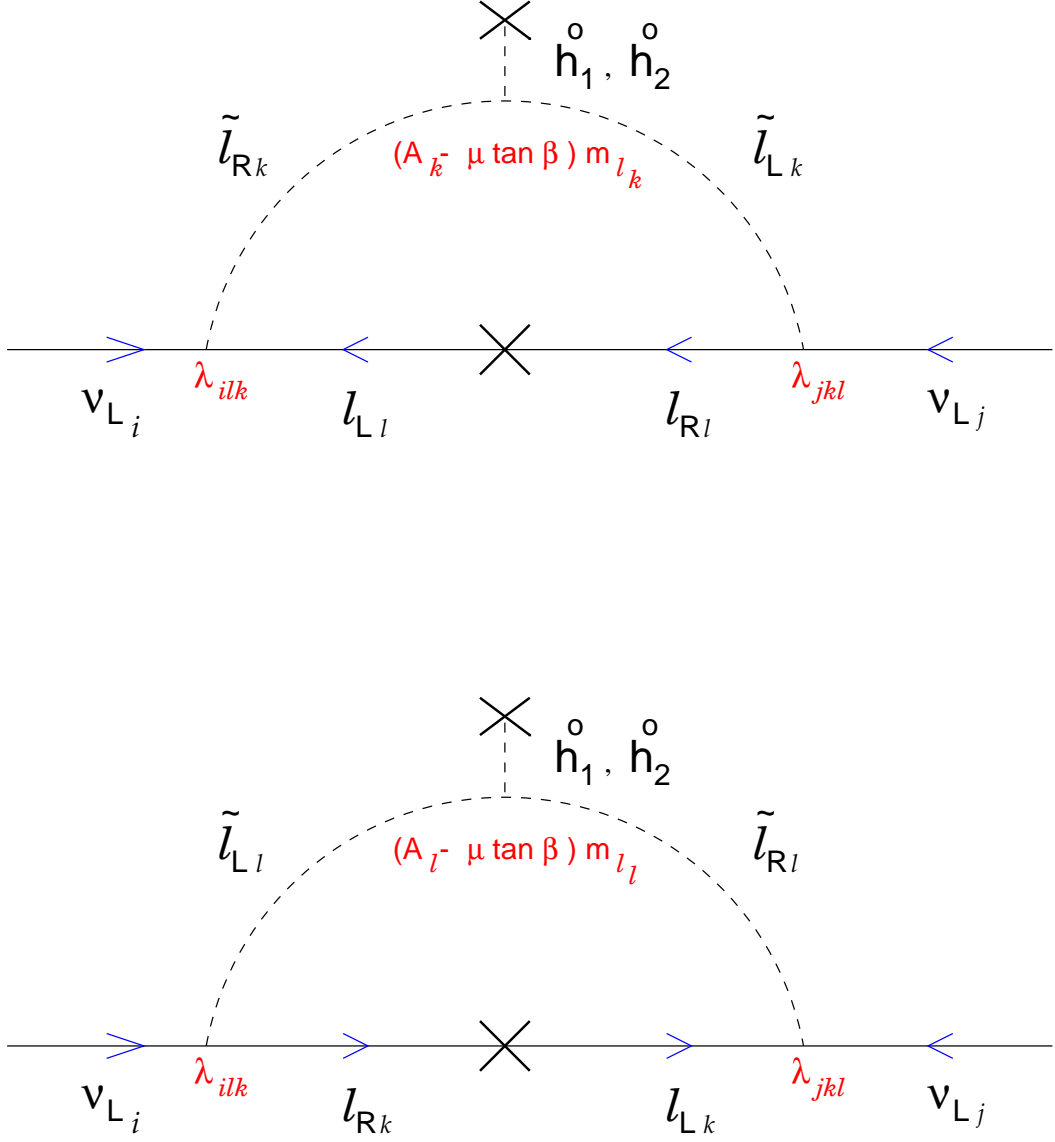}
\vskip1in
\caption{\label{fig2}
\small 
The one-loop diagrams for neutrino mass generated by the usual term
$m_{\scriptscriptstyle \ell}\; (A^{\!\scriptscriptstyle E} - \mu \tan\beta)$ 
in $LR$ slepton mixing. }
\end{figure}

\begin{figure}[th]
\centering
\includegraphics[width=5.5in]{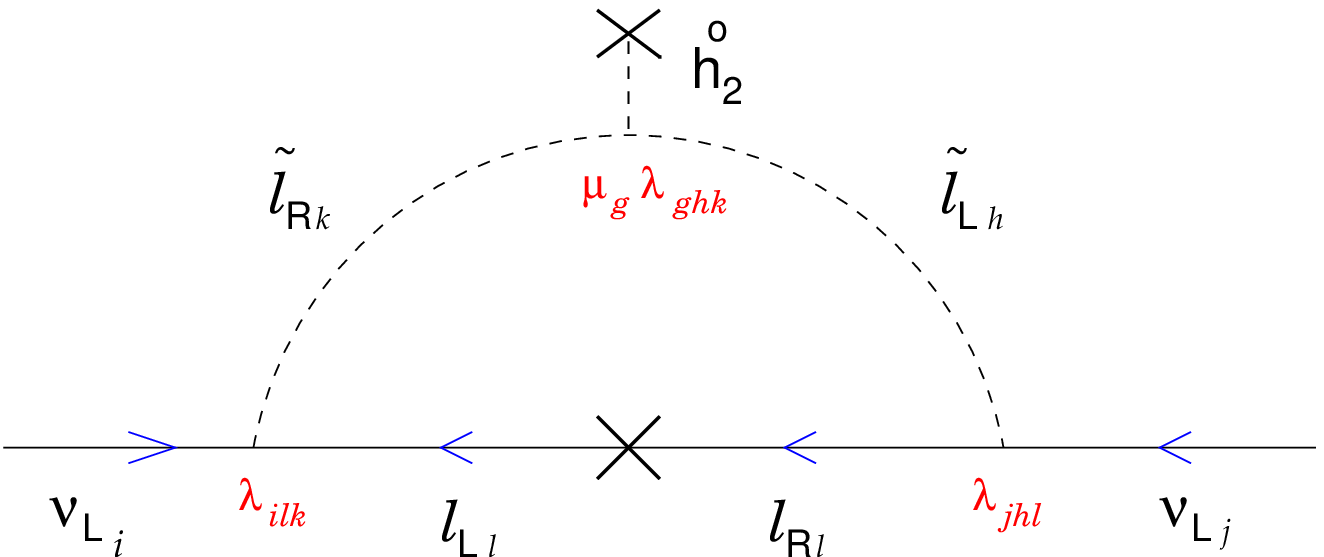}
\vskip1in
\caption{\label{fig3}
\small 
The one-loop diagram for neutrino mass generated by the term 
$\langle h_{\scriptscriptstyle 2}^0 \rangle\, \mu_g \lambda_{ghk}\, 
\tilde{\ell}_{\scriptscriptstyle L_h} \tilde{\ell}^*_{\scriptscriptstyle R_k}$
in the $LR$ slepton mixing.
}
\end{figure}

As mentioned in the Introduction, an easy way to complete
the Zee diagram without the $\mu_i$-type, bilinear $R$-parity-violating, 
couplings is to introduce an additional pair of Higgs doublet superfields. 
Denoting 
them by $H_{\!\scriptscriptstyle 3}$ and $H_{\!\scriptscriptstyle 4}$, 
bearing the same quantum numbers as $H_{\!\scriptscriptstyle 1}$ and 
$H_{\!\scriptscriptstyle 2}$, respectively, $R$-parity-violating terms
of the form
\[
\epsilon_{\!\scriptscriptstyle \alpha\beta} \lambda^{\!\scriptscriptstyle H}_k
H_{\!\scriptscriptstyle 1}^\alpha H_{\!\scriptscriptstyle 3}^\beta
E_k^c 
\]
can be introduced. With a trivial extension of notations (in Fig. ~\ref{fig1}
with $h_2^0$ replaced by $h_3^0$), we obtain 
a Zee diagram contribution to $(m_\nu)_{ij}$ through $\lambda_{ijk}$ as 
follows :
\begin{equation} \label{h1i}
\frac{-1}{16\pi^2} \, \frac{\langle h_{\scriptscriptstyle 3}^0 \rangle} 
{\langle h_{\scriptscriptstyle 1}^0 \rangle}\;  
(m_{\scriptscriptstyle \ell_j}^2 -m_{\scriptscriptstyle \ell_i}^2) \;
\lambda_{ijk}\;
\lambda^{\!\scriptscriptstyle H}_k A^{\!\scriptscriptstyle H}_k  \;
f( M_{h_{\scriptscriptstyle 1}^{\mbox{-}}}^2, M_{\tilde{\ell}_{R_k}}^2 ) \;.
\end{equation}
Here the slepton $\tilde{\ell}_{\!\scriptscriptstyle R_k}$ keeps the role
of the Zee scalar. We have neglected  the $F$ term obtainable in the
existence of bilinear $\mu$ terms between $H_{\!\scriptscriptstyle 3}$
and $H_{\!\scriptscriptstyle 2}$ or $H_{\!\scriptscriptstyle 4}$. At least
one of them has to be there. 
When no $L_i H_{\!\scriptscriptstyle 3} E^c_j$ type 
$R$-parity-conserving Yukawa couplings are allowed, the only surviving
extra contribution to neutrino mass among those discussed
is from the one corresponding to expression (\ref{Aloop}). Notice that
the second Higgs doublet of the Zee model, corresponding to 
$H_{\!\scriptscriptstyle 3}$ here, is also assumed not to have couplings
of the form $L_i H_{\!\scriptscriptstyle 3} E^c_j$. The condition for this 
$LR$ slepton mixing contribution to be below the required $m_{e\mu}$
would be the same as discussed in the previous section. 

However, there is a new contribution to $(m_\nu)_{kk}$ given by
\begin{equation} \label{hkk}
\frac{-1}{16\pi^2} \, \frac{\langle h_{\scriptscriptstyle 3}^0 \rangle\!^2} 
{ \langle h_{\scriptscriptstyle 1}^0 \rangle\!^2}\; 
m_{\!\scriptscriptstyle \ell_k}^2
(\lambda^{\!\scriptscriptstyle H}_k)^2 A^{\!\scriptscriptstyle H}_k  \;
f( M_{h_{\scriptscriptstyle 1}^{\mbox{-}}}^2, M_{\tilde{\ell}_{R_k}}^2 ) \;,
\end{equation}
which is obtained from Fig.~\ref{fig1}
with $\ell_{\scriptscriptstyle L}$ replaced by 
$\tilde{h}^{\mbox{-}}_{\scriptscriptstyle 1}$ and $h^0_{\scriptscriptstyle 2}$
by $h^0_{\scriptscriptstyle 3}$ (of course with different couplings at the
vertices.)  
This is in fact a consequence of the fact that the 
term $\lambda^{\!\scriptscriptstyle H}_k
H_{\!\scriptscriptstyle 1}^\alpha H_{\!\scriptscriptstyle 3}^\beta
E_k^c$ provides new mass mixings for the charged 
Higgsinos and the charged leptons.
As with the mixings induced by the $\mu_i$'s, the new effect is 
see-saw suppressed; but unlike the $\mu_i$'s their magnitude may be less 
severely constrained.  Nevertheless, the
essential difference here is that unlike the $\mu_i$ terms the 
$\lambda^{\!\scriptscriptstyle H}_k 
H_{\!\scriptscriptstyle 1}^\alpha H_{\!\scriptscriptstyle 3}^\beta E_k^c$ 
term does not contribute to the
mixings between neutrinos and the gauginos and higgsinos on tree level.

We will assume also that the deviations of the charged-lepton mass eigenstates
resulted from the new mixing are negligible. Bleaching the assumption actually 
does not cause too much trouble though. Its main effect is simply the 
modification of the numerical values of $m_{\!\scriptscriptstyle \ell_i}$'s 
used as the physical
masses get extra contributions. Here, similar to the above
we are interested in only the minimal set of couplings 
$\{\lambda_{{\scriptscriptstyle 12}\,k}\;,\;
\lambda_{{\scriptscriptstyle 13}\,k}\;,\; 
\lambda^{\!\scriptscriptstyle H}_{\scriptscriptstyle k}\}$ with a specific $k$.
For expression (\ref{h1i}) to give the right value to $m_{e\mu}$, we need
\begin{equation}
\lambda_{{\scriptscriptstyle 12}\,k}\, \lambda^{\!\scriptscriptstyle H}_k
\; \sim \frac{\mbox{Max}( M_{h_{\scriptscriptstyle 1}^{\mbox{-}}}^2, 
M_{\tilde{\ell}_{R_k}}^2 )}{A^{\!\scriptscriptstyle H}_k}\;
\frac{\langle h_{\scriptscriptstyle 1}^0 \rangle}
{\langle h_{\scriptscriptstyle 3}^0 \rangle} 
\; (7\times 10^{-7} \,\mbox{GeV}^{-1})\;, 
\end{equation}
and similarly for $m_{e\tau}$, it requires
$\lambda_{{\scriptscriptstyle 13}\,k}= (m_\mu^2/m_\tau^2)
\lambda_{{\scriptscriptstyle 12}\,k}\;$. This condition is easy to satisfy,
for example, when we take  $\langle h_{\scriptscriptstyle 3}^0 \rangle /
 \langle h_{\scriptscriptstyle 1}^0 \rangle  = 0.1$. 
Next, we compare the expression (\ref{h1i}) with Eq. (\ref{hkk}). 
For Eq. (\ref{h1i}) to dominate over Eq. (\ref{hkk}), it is required that
\begin{eqnarray}
\lambda_{{\scriptscriptstyle 12}\,k} &\gg& \lambda^{\!\scriptscriptstyle H}_k
\; \frac{\langle h_{\scriptscriptstyle 3}^0 \rangle} 
{\langle h_{\scriptscriptstyle 1}^0 \rangle}\; 
\frac{m^2_{\scriptscriptstyle \ell_k}}{m^2_\mu} \;,
\nonumber \\
\lambda_{{\scriptscriptstyle 13}\,k} &\gg& \lambda^{\!\scriptscriptstyle H}_k
\; \frac{\langle h_{\scriptscriptstyle 3}^0 \rangle} 
{\langle h_{\scriptscriptstyle 1}^0 \rangle}\;
 \frac{m^2_{\scriptscriptstyle \ell_k}}{m^2_\tau} \;.
\end{eqnarray}
The most favorable scenario under the context is obtained by taking $k=1$ 
where $m_{\scriptscriptstyle \ell_k}$ is just the $m_e$. The above requirements
are then easily satisfied. In addition, the corresponding requirement for
subdomination of the $LR$ slepton mixing contribution discussed above is then 
the same as Eq. (\ref{lamll}), and we also have Eq. (\ref{mudy}) 
from the tree-level Zee-scalar induced muon decay. All these constraints
can now be easily satisfied. Hence, having such a supersymmetric Zee model
looks very feasible.

One may argue, following the spirit of the single-VEV parametrization, that
$H_{\!\scriptscriptstyle 3}$ may be arranged to have no VEV. That would 
apparently kill the scenario. However, we have an assumption
above that there is no $L_i H_{\!\scriptscriptstyle 3} E^c_j$ term in the
superpotential, which is also adopted in the original Zee model.
Without the assumption we could then switch $H_{\!\scriptscriptstyle 3}$ and
$H_{\!\scriptscriptstyle 1}$ around, and though the scenario is still viable
it, however,  becomes much more complicated to analyze. 
Furthermore, if $H_{\!\scriptscriptstyle 1}$ was the only one with a VEV,
$h{\scriptscriptstyle 3}^{\mbox{-}}$ should take over the role of 
$h{\scriptscriptstyle 1}^{\mbox{-}}$ in Fig.~\ref{fig1}, and
then the couplings of the $h{\scriptscriptstyle 3}^{\mbox{-}}$ to leptons 
could not be taken diagonal in general. 
Studies of these more general situations, together with more admissible 
terms in the superpotential involving $H_{\!\scriptscriptstyle 3}$, 
actually worth more attention.  
This is however beyond the scope of the present paper.

An alternative approach is to give up identifying the right-handed slepton 
as the Zee scalar. One can introduce a 
vectorlike pair of Zee (singlet) superfields $E_{\!\scriptscriptstyle Z}$ and 
$E_{\!\scriptscriptstyle Z}^c$ with the scalar component of the latter
as the Zee scalar. A 
$\lambda_{ij}^{\!\scriptscriptstyle Z} L_i L_j E_{\!\scriptscriptstyle Z}^c$ 
term takes the role 
of the $\lambda_{ijk}$ above.  The $F$ term of $L_k$ with nonzero 
$\mu_k L_k H_{\!\scriptscriptstyle 2}$ and 
$Y_k^{\!\scriptscriptstyle Z} L_k H_{\!\scriptscriptstyle 1} 
E_{\!\scriptscriptstyle Z}^c$ terms
provides the mixing between the new Zee scalar and the 
$h_{\scriptscriptstyle 1}^{\mbox{-}}$. 
But the $Y_k^{\!\scriptscriptstyle Z}$ coupling 
easily messes up the identity of the physical 
charged leptons. It is clear then this is an even more complicated situation
than the previous one, and has to be analyzed carefully in a different 
framework.

Finally, one can take the trivial
supersymmetrization by taking both $E_{\!\scriptscriptstyle Z}$ and 
$E_{\!\scriptscriptstyle Z}^c$ as well as $H_{\!\scriptscriptstyle 3}$ and 
$H_{\!\scriptscriptstyle 4}$. 
The restrictions on the parameter space of the relevant couplings are 
then unlikely to have any interesting feature beyond that of the
Zee model itself. It is interesting, however, to note that the couplings
needed, $L_i L_j E_{\!\scriptscriptstyle Z}^c$ and 
$H_{\!\scriptscriptstyle 1} H_{\!\scriptscriptstyle 3} E_{\!\scriptscriptstyle
 Z}^c$,
do not break $R$ parity at all, though the lepton number is violated. 

\section{conclusions}
We have discussed the embedding of the Zee neutrino mass model in the 
framework of $R$-parity-violating supersymmetry. It is a nontrivial
supersymmetrization of the Zee model in the sense that one or both of
the extra scalar fields of the Zee model are identified within the
minimal supersymmetric SM spectrum. We have analyzed in detail the minimal
scheme where a right-handed slepton plays the role of the charged Zee
scalar, with no extra Higgs doublet introduced.  
The $\lambda_{ijk}$ and $\mu_i$ couplings in $R$-parity-violating supersymmetry
are identified as the lepton-number-violating couplings in the Zee model. 
Nevertheless, the scheme also introduces other types of contributions to 
neutrino masses.  We have described in detail all these contributions and
their general forms.

We have addressed and answered the question on the feasibility of such a
model. For the minimal scheme we illustrated that a set of 
$R$-parity-violating couplings given as $\{\lambda_{{\scriptscriptstyle
12}\,k}\;,\; \lambda_{{\scriptscriptstyle 13}\,k}\;,\;
\mu_{\scriptscriptstyle k}\}$ with a specific $k$ completes the model.
The various contributions to neutrino masses are discussed and the
conditions for maintaining the zeroth order Zee mass texture
and for fitting the experimental neutrino oscillation data
are derived.  The case with $k=3$ has been shown to be marginally feasible,
though the constraints are very stringent.  Here, the right-handed
stau is the Zee scalar. The analysis also illustrates the interesting
interplay between different couplings, and the relative strength of
each type of contributions.  
In particular, we have discussed two contributions both involving
the bilinear $R$-parity violating couplings $\mu_{\scriptscriptstyle k}$, 
and the trilinear ones.  Both contributions have not been discussed before
and the Zee mechanism does correspond to one of them and the another one 
involves a new source of $LR$ slepton mixing.

We have also discussed alternatives to the minimal scheme.
Among the alternatives, we have considered the interesting case with an extra
pair of vectorlike Higgs superfields. A
$\lambda^{\!\scriptscriptstyle H}_{\scriptscriptstyle k}$ coupling
for the superpotential term $H_{\!\scriptscriptstyle 1}^\alpha
H_{\!\scriptscriptstyle 3}^\beta E_k^c $ replaces
$\mu_{\scriptscriptstyle k}$ of the previous case. Constraints on this
extended type of models are much weaker, and so having a phenomenologically 
viable model of this type would not be a problem. The best scenario in this 
case 
is for $k=1$, namely, taking the selectron as the Zee scalar. More detailed
studies of such models worth a serious effort. We hope to report on
that in the future.

\acknowledgements
This research was supported in part by the U.S.~Department of Energy under
Grants No. DE-FG03-91ER40674 and by the Davis Institute for High Energy 
Physics.
OK wants to thank Darwin Chang for discussions, and colleagues at Academia 
Sinica, especially Hai Yang Cheng, for support.


\clearpage

\begin{thebibliography}{99}
\bibitem{mueg}
Note that a major source of constraint from $\mu \to e \gamma$ in the general
Zee model scenario is absent here as we have $f_{\scriptscriptstyle 23}=0$.
\end{thebibliography}
\end{document}